# Extended-variable probabilistic computing with p-dits

Christian Duffee[1*], Jordan Athas[1], Andrea Grimaldi[2],

Deborah Volpe[3], Giovanni Finocchio[4*], Ermin Wei[1,5], Pedram Khalili Amiri[1,6*]

**Affiliations**

[1] Department of Electrical and Computer Engineering, Northwestern University, Evanston, IL 60208, United States of America.

[2] Department of Electrical and Information Engineering, Politecnico di Bari, Bari, 70126, Italy.

[3] Istituto Nazionale di Geofisica e Vulcanologia, Rome, 00143, Italy.

[4] Dipartimento di Scienze Matematiche e Informatiche, Scienze Fisiche e Scienze della Terra, Università degli Studi di Messina, Messina, 98166, Italy

[5] Department of Industrial Engineering and Management Sciences, Northwestern University, Evanston, IL 60208, United States of America.

[6] Applied Physics Program, Northwestern University, Evanston, IL 60208, United States of America.

∗ Emails: christian.duffee@northwestern.edu; gfinocchio@unime.it; pedram@northwestern.edu

**Abstract**

Ising machines can solve combinatorial optimization problems by representing them as energy minimization problems. A common implementation is the probabilistic Ising machine (PIM), which uses probabilistic (p-) bits to represent coupled binary spins. However, many real-world problems have complex data representations that do not map naturally into a binary encoding, leading to a significant increase in hardware resources and time-to-solution. Here, we describe a generalized spin model that supports an arbitrary number of spin dimensions, each with an arbitrary real component. We define the probabilistic d-dimensional bit (p-dit) as the base unit of a p-computing implementation of this model. We further describe two restricted forms of p-dits for specific classes of common problems and implement them experimentally on an application-specific integrated circuit (ASIC): (A) isotropic p-dits, which simplify the implementation of categorical variables resulting in ~34x performance improvement compared to a p-bit implementation on an example 3-partition problem. (B) Probabilistic integers (p-ints), which simplify the representation of numeric values and provide ~5x improvement compared to a p-bit implementation of an example integer linear programming (ILP) problem. Additionally, we report a field-programmable gate array (FPGA) p-int-based integer quadratic programming (IQP) solver which shows ~64x faster time-to-solution compared to the best of a series of state-of-the-art software solvers. The generalized formulation of probabilistic variables presented here provides a path to solving large-scale optimization problems on various hardware platforms including digital CMOS.

**Introduction**

Many problems in computing, including important combinatorial optimization tasks, cannot be effectively solved by deterministic von Neumann computers[1,2]. Physics-inspired unconventional computing methodologies offer a potential path to solving such problems using fewer resources[3–8]. A sub-set of these approaches, collectively termed Ising machines, are inspired by the evolution of interacting spins in a magnetic material[9–12]. Variables in a problem are encoded as a collection of binary values, analogous to the up or down states of an electron spin. Coupling interactions between these spins create an energy landscape that can be formalized into a quadratic Hamiltonian, with the problem's solution corresponding to its energy ground state.

While Ising machines can be simulated using conventional processors[13], a wide range of physical phenomena have been explored to develop dedicated hardware platforms for their realization. Examples include systems of interacting magnetic[14–16], optical[17–21] and electronic complementary metal-oxide semiconductor (CMOS) oscillators[22–27], as well as polariton based systems[28–30]. Here we will focus on a promising class of hardware implementations known as probabilistic Ising machines (PIMs)[7,31–40]. A PIM is composed of probabilistic bits (p-bits) that stochastically switch between a $+1$ and $-1$ state. During its operation, weighted probabilistic interactions among its p-bits have a net effect of driving the PIM towards an energy minimum, where the desired solution resides. To improve the likelihood and speed of finding the ground configuration, sophisticated energy minimization techniques are often employed. For instance, a decreasing mathematical temperature, termed simulated annealing, a swapping of configurations between parallel problem instances of differing temperatures, termed parallel tempering, or a decreasing mathematical quantum-inspired interaction between parallel problem instances, termed simulated quantum annealing, are used to assist in finding a solution[41–44].

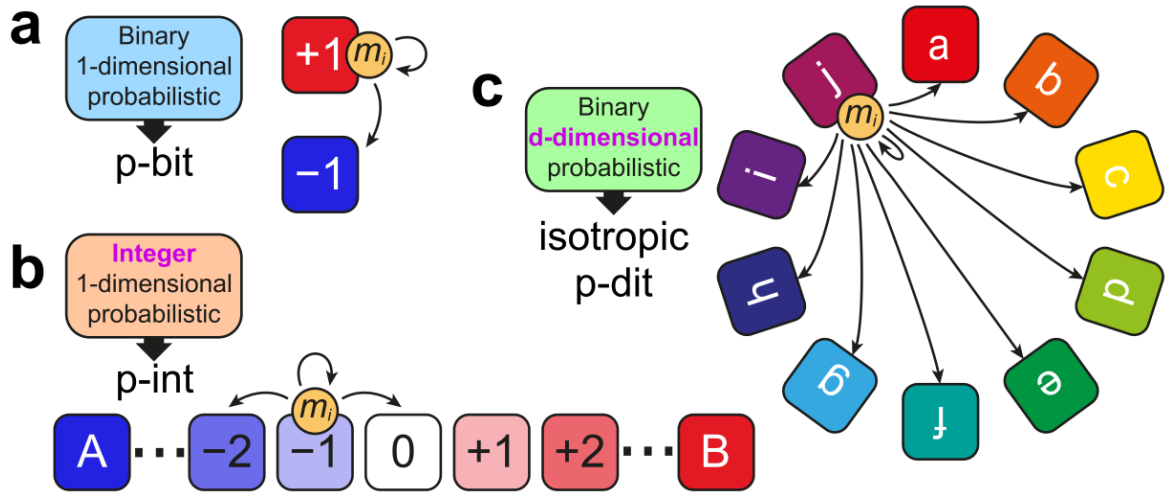

**Figure 1: Overview of extended probabilistic variables.** The general p-dit described here reduces to a p-bit, p-int, or isotropic p-dit, depending on the number of dimensions and range of values that are allowed. **A** A p-bit is a binary, 1-dimentional element which switches between one of two opposing states. **B** A p-int has a larger range of possible values compared to a p-bit. It moves between integer states which exist along a single dimension. These states are bounded on the lower end by the constant $A$ and on the upper end by the constant $B$. **C** An isotropic p-dit can be considered as a p-bit with extra dimensions. It switches between orthogonal states, each of which exists along a separate dimension. The example shown here is an isotropic p-dit with 10 dimensions, hereafter denoted as an isotropic p-dit[10].

While many problems map naturally to this representation by using a single p-bit per variable, others do not, requiring either multiple p-bits per variable or auxiliary p-bits (which do not directly represent any variable) to be implemented in an Ising model[32,45]. This expands the state space and introduces additional local energy minima, which the PIM can falsely converge to[46]. To mitigate this issue, here we theoretically describe a continuous generalized spin model, where each spin can hold an arbitrary component length in $|D|$-dimensional space. We further formalize the concept of a probabilistic d-dimensional bit (p-dit) which stochastically oscillates between a set of discrete points in this space. We then describe and experimentally demonstrate two practical restrictions on these states, illustrated in Fig. 1: (A) isotropic p-dits and (B) probabilistic integers (p-ints).

The former, isotropic p-dits, are units that probabilistically oscillate between several orthogonal states based on their energy levels. Isotropic p-dits are designed to implement a categorical variable's value, which is conventionally represented with a group of p-bits using one-hot encoding[45]. Importantly, one-hot encodings are invalid if either zero, or more than a single p-bit are in the +1 state within each group. While less important for small problems, the ratio of invalid to valid possible one-hot assignments grows exponentially with the number of variables, as demonstrated in Fig. S1. In contrast, an isotropic p-dit-based categorical assignment will not have any invalid states, greatly reducing the size of the energy landscape that must be explored. While many problems can take advantage of isotropic p-dits, we demonstrate their strength using the *N*-partition problem.

The latter, p-ints, hold an integer value within a predefined range. P-ints can replace Ising representations where collections of p-bits are used to represent an integer value under a base-2 representation. Compared to collections of p-bits, p-ints have reduced interaction complexity and allow smoother (in terms of energy barriers) transitions within the state space between adjacent values. We experimentally demonstrate their applicability and advantage in selected integer linear programming (ILP) and integer quadratic programming (IQP) problems.

To showcase the effectiveness of these extended variables, we designed an application-specific integrated circuit (ASIC) that integrates p-bits, p-ints, and isotropic p-dits using a 130 nm foundry process. The ASIC, which is controlled by an embedded RISC-V central processing unit (CPU), contains all elements needed to implement general probabilistic computing: (i) programable weight and offset matrix, (ii) nonlinear (sigmoid) probability curve implementation, (iii) tunable temperature for simulated annealing, and (iv) several tunable hyper-parameters.

Despite constraints on chip area and technology node, the ASIC illustrates the feasibility of implementing extended p-computing variables in a scalable digital CMOS platform and allows us to experimentally demonstrate their superiority in solving representative three-partition and ILP problems.

**Model Formulation**

*Generalized Spin Model*

Let us define a set of bases in a $|D|$-dimensional space as

$$D = \{\hat{a}, \hat{b}, \hat{c}, ...\}, |D| \geq 2. \tag{1}$$

Let any collection of spins be ordered from 1 to $N$ without loss of generality. Define the $i^{\text{th}}$ spin by its real components along each of $|D|$ bases,

$$m_i = \begin{bmatrix} m_i^1 \\ m_i^2 \\ \vdots \\ m_i^{|D|} \end{bmatrix}, \forall_a m_i^a \in \mathbb{R}. \tag{2}$$

Let a bias matrix of size $N \times |D|$ with component

$$h_i = \begin{bmatrix} h_i^1 \\ h_i^2 \\ \vdots \\ h_i^{|D|} \end{bmatrix}, \forall_a h_i^a \in \mathbb{R}, \tag{3}$$

representing the bias on the $i$th spin exist, such that $-m_i^T h_i$ represents the first order effect of the $i^{\text{th}}$ spin on the energy of the many-spin system. Similarly, let a coupling tensor of size $N \times N \times |D| \times |D|$ with component

$$J_{ij} = \begin{bmatrix} J_{ij}^{11} & \cdots & J_{ij}^{1|D|} \\ \vdots & \ddots & \vdots \\ J_{ij}^{|D|1} & \cdots & J_{ij}^{|D||D|} \end{bmatrix}, \forall_{a,b} J_{ij}^{ab} \in \mathbb{R}, \tag{4}$$

representing the coupling effect of the $j^{th}$ spin on the $i^{th}$ exist, such that $-\frac{1}{2}m_i^T J_{ij} m_j$ corresponds to the second order effect on the energy of the system. The total energy of a system of $N$ spins with second order interactions then is the quadratic Hamiltonian,

$$E_S = -\left(\sum_{i=1}^{N} m_i^T h_i + \frac{1}{2}\sum_{i=1}^{N}\sum_{j=1}^{N} m_i^T J_{ij} m_j\right). \tag{5}$$

In general, higher order interactions can be defined as a series of coupling tensors, $T_{ijk...}$ of various sizes such that, using explicit indexing,

$$E_S = -\left(\frac{1}{1!}\sum_{i=1}^{N}\sum_{a=1}^{|D|} m_i^a h_i^a + \frac{1}{2!}\sum_{i=1}^{N}\sum_{j=1}^{N}\sum_{a=1}^{|D|}\sum_{b=1}^{|D|} m_i^a m_j^b m_k^c J_{ij}^{ab} \right. \tag{6}$$

$$\left. + \frac{1}{3!}\sum_{i=1}^{N}\sum_{j=1}^{N}\sum_{k=1}^{N}\sum_{a=1}^{|D|}\sum_{b=1}^{|D|}\sum_{c=1}^{|D|} m_i^a m_j^b m_k^c T_{ijk}^{abc} + \cdots\right).$$

The influence acting upon on the $i$th spin along the $a$th dimension can be defined as

$$I_i^a = h_i^a + \sum_j^{N}\sum_b^{|D|} J_{ij}^{ab} m_j^b + \sum_j^{N}\sum_k^{N}\sum_b^{|D|}\sum_c^{|D|} T_{ijk}^{abc} + \cdots. \tag{7}$$

We often wish to assume symmetric coupling both in terms of dimension and spins; for a second order system, this requires $J^{ab} = J^{ba}$ and $J_{ij} = J_{ji}$. The energy difference between the $i$th spin of a second-order system being in two states is then

$$\Delta E_i^{m \to m^*} = -\left[\sum_{a=1}^{|D|} m_i^{a*}\left(I_i^a - \frac{1}{2}\sum_{b=1}^{|D|} J_{ii}^{ab}(2m_i^b - m_i^{b*})\right) - \sum_{a=1}^{|D|} m_i^a\left(I_i^a - \frac{1}{2}\sum_{b=1}^{|D|} J_{ii}^{ab} m_i^b\right)\right]. \tag{8}$$

*Probabilistic Computing with P-dits*

To perform probabilistic computing by using this generalized spin model, let us define a probabilistic d-dimensional bit (p-dit) which oscillates between discrete spin values in the set $M_i$. If a single p-dit updates at a time and considers all states, including its current state, then the probability of an arbitrary state, using Glauber-like dynamics, is

$$P_i^{x \in |M_i|} = \frac{1}{1 + \sum_{y \in |M_i| \setminus x} \exp(\beta \Delta E_i^{y \to x})} = \sigma_{|M_i|-1}(-\beta \Delta E_i^{y_1 \to x}, -\beta \Delta E_i^{y_2 \to x}, \ldots), \quad (9)$$

where $\sigma_x$ is a generalized sigmoid function with $x$ inputs. It is shown in Supplementary Note 1, that this results in a system configuration probability distribution that converges to exactly the Boltzmann distribution,

$$\pi_{S_i} = \frac{\exp(-\beta E_{S_i})}{\sum_j^W \exp(-\beta E_{S_j})}, \quad (10)$$

if there are a total of $W$ possible system configurations ($S_i$). Alternatively, only a limited number of states, based on the current state, can be considered. If these, for the current state indexed $a \in |M_i|$, are $G_i^a \subseteq M_i$, then the update probability is

$$P_i^{x \in |G_i^a|} = \frac{1}{1 + \sum_{y \in |G_i^a| \setminus x} \exp(\beta \Delta E_i^{y \to x})} = \sigma_{|G_i^a|-1}(-\beta \Delta E_i^{y_1 \to x}, -\beta \Delta E_i^{y_2 \to x}, \ldots), \quad (11)$$

which is shown in Supplementary Note 2 to converge if all states are reachable from each other. This is sometimes preferable to increase the speed at which each update can be computed, or to encourage collections of p-dits to explore configurations that are close together.

Relation to the XY and Planar/Clock Potts Model

Note that the XY model[29] can be interpreted as a subset of the generalized model. Specifically, for an XY model with coupling matrix $J'$ and bias field vector $h'$, we can restrict the generalized model to

$$|D| = 2$$

$$|m_i| = 1$$

$$\forall_{i,a} l_i^a \geq 0 \tag{12}$$

$$h_i = \begin{bmatrix} h_i' \\ 0 \end{bmatrix}$$

$$J_{ij} = \begin{cases} 0, & i = j \\ \begin{bmatrix} J_{ij}' & 0 \\ 0 & J_{ij}' \end{bmatrix}, & i \neq j. \end{cases}$$

This XY model can be expanded by replacing the zero elements in $h_i$ and $J_{ij}$, for $i \neq j$, to incorporate arbitrary angle, rather than just 180°, bias and coupling.

With these restrictions, we can define that each $M_i$ contains $q \geq 2$ possible states that are evenly distributed on the unit circle. Adjacent states are separated by an angle equal to

$$\theta = \frac{2\pi}{q}. \tag{13}$$

This is the discrete version of the model referred to as the planar Potts model or clock Potts model[28]. Notably, for $q \geq 4$, a given state will have two neighboring states, with all others separated by a greater angle. As such, independent coupling between pairs of states is not possible, since a planar field along any one state's angle imparts a greater stray field component on the state's neighbors compared to its non-neighbors.

### *Probabilistic Computing with P-bits*

A traditional p-bit PIM is designed to find the energy minimum of the Ising Hamiltonian formulation of computing problems with each p-bits having the state, $m$, of either $-1$ or $+1$. Notably, there are no self-coupled p-bits and all coupling interactions are symmetric[47–49]. As such, we can represent p-bits as a p-dit with the restrictions that

$$|D| = 1$$

$$M_i = \{[+1], [-1]\} \tag{14}$$

$$J_{ii}^{11} = J_{ii} = 0.$$

In doing so, we return to the traditional p-bit[47–50] Hamiltonian,

$$E_S = -\left(\sum_{i=1}^{N} h_i m_i + \sum_{i=1}^{N}\sum_{j=1}^{N} \frac{1}{2} J_{ij}\, m_i m_j\right), \tag{15}$$

energy difference,

$$\Delta E_i^{-1\uparrow 2} = E_i^{+1} - E_i^{-1} = -2I_i, \qquad I_i = h_i + \sum_{j}^{N} J_{ij}\, m_j, \tag{16}$$

and update probability,

$$P_i^{+1} = \frac{1}{1 + \exp(\beta \Delta E_i^{-1\uparrow 2})} = \sigma(-\beta \Delta E_i^{-1\uparrow 2}). \tag{17}$$

An expanded derivation of Equation (16) and Equation (17) is shown in Supplementary Equation 1 and Supplementary Equation 2, respectively.

A p-int can be considered as a generalization of this one-dimensional p-bit formulation, allowing more than two integer values along a single dimension. An alternate p-dit restriction where $|D| = 2$ and $M_i = \{[1\ 0]^T, [0\ 1]^T\}$, also acts identically to a p-bit and is discussed in Supplementary Note 3. An isotropic p-dit can be considered as a generalization of this latter case, and will be described in the next section.

### *Probabilistic Computing with Isotropic P-dits*

Let us restrict a p-dit to lie only along one of the dimensional basis vectors and have no self-coupling, that is

$$M_i = \{[1\ 0\ ...\ 0]^T, [0\ 1\ ...\ 0]^T, [0\ 0\ ...\ 1]^T\} \tag{18}$$

$$\forall_i J_{ii} = 0.$$

In this formulation, a p-dit naturally encodes a categorical variable with $|D|$ possible options.

For some problems, every variable option is analogous and does not have a meaning outside of which p-dits hold each value. Examples include many graph coloring and set partition problems where the labeling is arbitrary. In this special case, we can define an isotropic p-dit such that

$$J_{ij}^{ab} = \begin{cases} +J'_{ij}, & a = b \wedge i \neq j \\ -J'_{ij}, & a \neq b \wedge i \neq j \\ 0, & i = j \end{cases} \tag{19}$$

Notably, the coupling dynamics can be defined by a single $N \times N$ matrix, $J'$, easing hardware implementation.

As an example, a three-dimensional isotropic p-dit, or isotropic p-dit³, has three possible states: $a$, $b$ and $c$. If two isotropic p-dits³ were positively (i.e., ferromagnetically) coupled, and the first was in state $\hat{a}$, then it would impart a force on the second that would pull it towards state $\hat{a}$ and away from both states $\hat{b}$ and $\hat{c}$ when updated. In contrast, if they were negatively (i.e., antiferromagnetically) coupled, the second p-dit³ would be pushed away from $\hat{a}$ and towards both $\hat{b}$ and $\hat{c}$.

The equation for the system energy difference between the $i^{\text{th}}$ isotropic p-dit being in two states, indexed c and $d$ is,

$$\Delta E_i^{c \to d} = -(I_i^d - I_i^c), \tag{20}$$

if

$$I_i^a = h_i^a + \sum_{b}^{|D|} J_{ij}^{ab} m_j^b. \tag{21}$$

Consequently, the probability of an isotropic p-dit switching to some state $x$ is

$$P_i^{x \in |D|} = \frac{1}{1 + \sum_{d \in |D| \wedge d \neq x}[\exp(\beta \Delta E_i^{d \to x})]} \tag{22}$$

$$= \sigma_{|D|-1}\left(-\beta(I_i^x - I_i^a), -\beta\left(I_i^x - I_i^b\right), \ldots, -\beta(I_i^x - I_i^w), -\beta(I_i^x - I_i^y), \ldots\right),$$

using a $|D|-1$ dimensional sigmoid function derived from Equation (10) in Supplementary Equation 3. It can be proved, as demonstrated in Supplementary Note 4, that this results in a state probability distribution that converges to the Boltzmann distribution, Equation (10).

While other works have proposed similar Ising-like representations of categorical variables, they have required explicit calculation of the total energy of each configuration rather than the energy difference between states[51], decomposition of the problem into Ising sub-problems[52], the use of an update function that selects a post-update state with uniform probability rather than the energy of each state[53], or[53]. The first strategy requires increased computation to perform each state update, while the latter two do not fully consider the multi-dimensional aspects of the energy landscape, potentially resulting in slower solution times.

*Probabilistic Computing with P-ints*

To represent interacting integer variables, we can imagine a p-dit with one dimension and multiple levels, each separated by 1 from the previous. That is, with a lower bound $A_i$ and an upper bound $B_i$,

$$|D| = 1 \tag{23}$$

$$M_i = \{[A_i], [A_i + 1], [A_i + 2], \ldots, [B_i - 1], [B_i]\}.$$

Like an isotropic p-dit, these probabilistic integers (p-ints) can hold one of more than two states. However, while in an isotropic p-dit, all states are equally orthogonal to each other, in a p-int these states are integers lying on a numbered line (e.g., state 3 is more positive than state 2 and opposite state −3). This distinction is visualized in Fig. 1, with a p-int's possible states illustrated on the bottom of the image, and the isotropic p-dit's possible states illustrated on the right of the image.

For p-ints to move along the single dimension by $k$, we get an energy difference of,

$$\Delta E_i^{m \to (m+k)} = -k(I_i + \frac{1}{2}J_{ii}k), \quad (24)$$

if

$$I_i = h_i + \sum_{j}^{N} J_{ij} m_j. \quad (25)$$

Notably, $J_{ij}$ is a single $N \times N$ matrix, which allows for compact hardware implementation. Let us consider the update scheme, which we will use for the rest of this work, where a p-int update considers its current state, the current state increased by one, and the current state decreased by 1. The energy difference between the incremented state and the current state is

$$\Delta E_i^{m \to m+1} = \Delta E_i^{m\uparrow 1} = E_i^{m+1} - E_i^m$$

$$= -\left(h_i + \sum_{j=1}^{N} J_{ij} m_j + \frac{1}{2}J_{ii}\right) = -I_i - \frac{1}{2}J_{ii}. \quad (26)$$

Similarly, comparing between the state with $m_i = (m-1)$ and the state with $m_i = (m)$ is

$$\Delta E_i^{m \to m-1} = \Delta E_i^{m\downarrow 1} = h_i + \sum_{j=1}^{N} J_{ij}m_j - \frac{1}{2}J_{ii} = I_i - \frac{1}{2}J_{ii}. \quad (27)$$

Consequently, between state with $m_i = (m+1)$ and state with $m_i = (m-1)$, we have an energy difference equivalent to that of a p-bit jump in Equation (16), given by

$$\Delta E_i^{m-1 \to m+1} = E_i^{m+1} - E_i^{m-1} = \Delta E_i^{m-1\uparrow 2} = -2I_i, \quad (28)$$

with full derivations shown in Supplementary Equation 4.

Using a two-dimensional generalization of the sigmoid function

$$\sigma_2(x, y) = \frac{1}{1 + \exp(-x) + \exp(-y)}, \quad (29)$$

which arises naturally from Equation (10) as shown in Supplementary Equation 5, the probability of switching to each state becomes

$$P_i^{m\uparrow 1} = \sigma_2\left(-\beta \Delta E_i^{m\uparrow 1}, -\beta \Delta E_i^{(m-1)\uparrow 2}\right) = \sigma_2\left[\beta\left(I_i + \frac{1}{2}J_{ii}\right), \beta(2I_i)\right]$$

$$P_i^{m\uparrow 0} = P_i^{m\downarrow 0} = \sigma_2\left(+\beta \Delta E_i^{m^*\uparrow 1}, +\beta \Delta E_i^{m\downarrow 1}\right) = \sigma_2\left[\beta\left(-I_i - \frac{1}{2}J_{ii}\right), \beta\left(I_i - \frac{1}{2}J_{ii}\right)\right] \quad (30)$$

$$P_i^{m\downarrow 1} = \sigma_2\left(-\beta \Delta E_i^{m\downarrow 1}, -\beta \Delta E_i^{(m+1)\downarrow 2}\right) = \sigma_2\left[\beta\left(-I_i + \frac{1}{2}J_{ii}\right), \beta(-2I_i)\right].$$

It can be shown that this formulation will result in a converging probability distribution, as discussed in Supplementary Note 5.

Crucially, the change in value is either an increment or a decrement of its current state. This contrasts with a p-bit, whose updated value does not directly depend on its previous state. The self-connection term adds a preference to moving towards (for negative values) or away (for positive values) from the $m_i = 0$ state. A p-int with an input value of zero and no self-connection has an equal chance of performing each of its three possible updates.

A key advantage of p-ints over p-bits is that they naturally represent a range of numbers instead of representing them through binary encoding[54]. This allows a p-int to smoothly progress down an energy gradient. On the other hand, a series of p-bits must take a roundabout path that may be energetically unfavorable. For instance, a p-int moving from 3 to 4 takes one step, while a series of p-bits attempting the same maneuver might first travel through numeric state 1 and 0 before reaching 4 (i.e., binary 011 to 001 to 000 to 100). As we show in the Results section, this difference has the effect of an improved Time-To-Solution (TTS) in ILP problems for p-ints compared to p-bits.

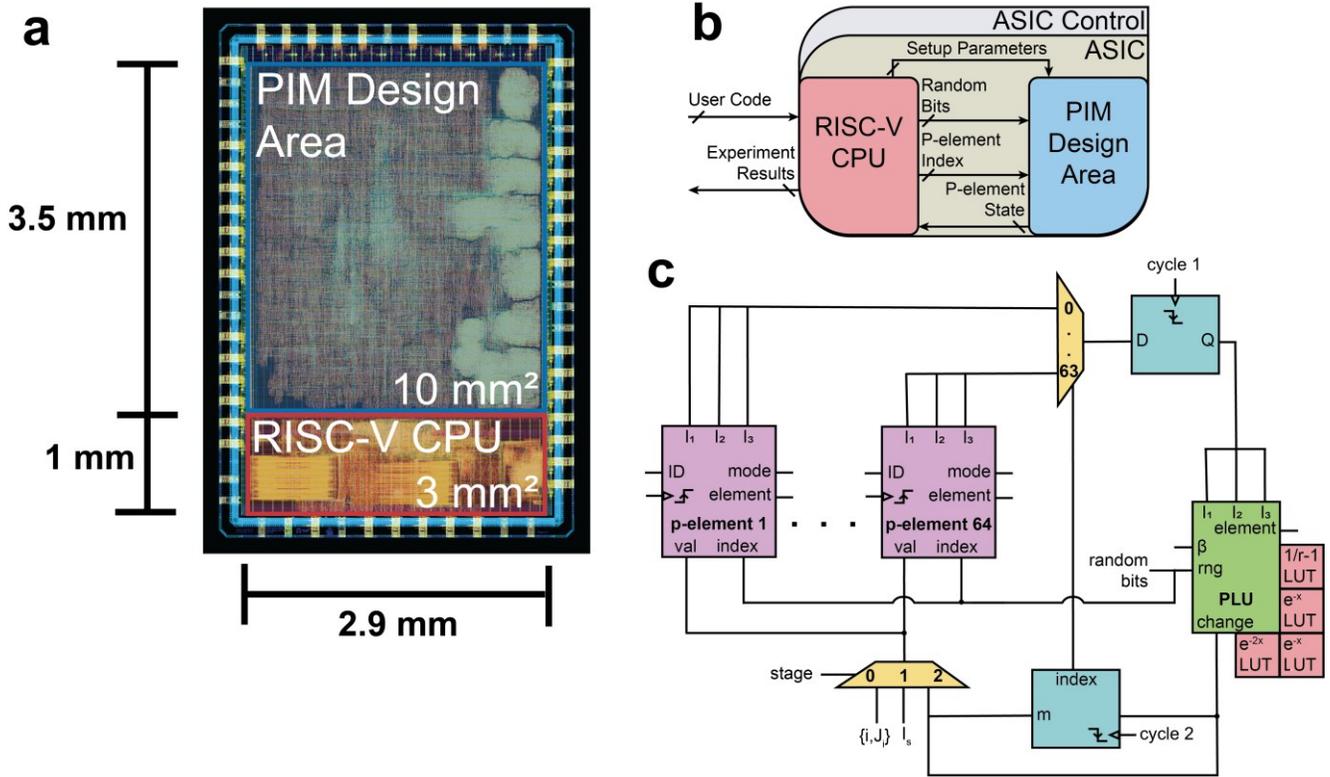

**Figure 2**: **P-computing hardware overview. A** Layout view of the manufactured p-computing ASIC. The chip is split between the PIM Design Area (top) and the driver RISC-V CPU (bottom). Surrounding it is circuitry that includes memory and I/O-related hardware. Metal layer 1 is excluded for visibility. **B** A simplified block diagram of the whole system. User code executed on the RISC-V driver CPU is used to initialize the PIM's starting state based on the chosen problem, provide random bits to the PIM during execution, and read out the final state of output p-elements. **C** A simplified block diagram of the PIM. During the first half of a two-clock period cycle, a random p-element is selected and its running $I$ totals are transmitted to the PLU, which computes the direction of its update. During the second half of the cycle, the identity of the updating p-element and the value of the update are broadcast over the 'index' and 'val' to all p-elements which, in turn, update their $I$ values. Each p-element has a unique identity index statically stored in the 'ID' signal. The interpretation of the $I$ values by both components vary depending on whether the 'element' signal is configured for p-bits, p-ints, or isotropic p-dits[3]. A multiplexer and operation 'mode' signal allow each p-element to be configured with their initial values.

**Probabilistic Computing Hardware Design**

To investigate the feasibility of implementing these extended probabilistic variables in hardware, and evaluate their advantage in specific optimization problems, an ASIC was designed and fabricated using the 130 nm Skywater foundry CMOS technology in an area of approximately 13 mm$^2$, as shown in Fig. 2a. This design contains 64 multi-purpose p-elements, with an all-to-all topology, each of which can act as either p-bits, p-ints, or isotropic p-dits[3]. The interactions are user-programmable via 70 bytes of memory within each p-element. Their updates are calculated by the probabilistic logic unit (PLU), which implements the necessary sigmoid transformation of their inputs, temperature scaling, and comparison to the random update threshold. A 10 MHz clock drives the system. A block diagram describing the working of the system as a whole and of the ASIC is shown in Fig. 2b and Fig. 2c, respectively.

C code running on a RISC-V CPU, which was built into the same chip, was used to control the PIM. To run each problem, first the CPU issues commands to the PIM to initialize the $J$ matrix, $h$ vector, the p-element mode, the p-element range bounds, and simulated annealing parameters. Afterwards, the PIM is moved into execution mode where a two-clock period update schedule is used. One clock period is used to compute the next state of a randomly selected p-element, and additionally update the used temperature, if a simulated annealing procedure is used. The alternate clock period is then used to update the internal state of all affected p-elements. Further implementation details are described in Supplementary Note 6.

**Encoding of Selected Problems**

*|D|-Partition*

The classical partition problem involves splitting a group of $N$ non-negative numbers into two subsets such that their sums are equivalent, or nearly equivalent. This problem has a natural Ising representation of

$$E = \sum_{i=1}^{N} \sum_{j \neq i}^{N} n_i n_j m_i m_j \therefore h = 0 \land J'_{ij} = \begin{cases} -2n_i n_j, & \text{if } i \neq j \\ 0, & \text{otherwise} \end{cases} \tag{31}$$

which can be implemented using p-bits. In this case, the state of each p-bit corresponds directly to the partition in which it exists.

A generalization of this is the $|D|$-dimension number partition problem, in which there are $|D|$ subsets rather than just two[45,48]. In this case, there is no one-to-one correspondence between single p-bit states and a number's subset. Instead, a collection of p-bits is needed for each number. One possible encoding is a one-hot representation where each number's subset placement is determined by $|D|$ p-bits. In addition to the anti-correlation between numbers required for the 2-partition case, this necessitates a constraint that exactly one of the $|D|$ p-bits for each number occupy the $+1$ state. Indexing the p-bits by number and partitioning them for readability yields the following Hamiltonian representation, with $E_C$ enforcing constraints and $E_O$ enforcing the objective,

$$\begin{aligned} E = E_C + E_O = C \sum_{i=1}^{N} \sum_{d=1}^{|D|} \left[ (|D| - 2) m_{i,d} + \sum_{e \neq d}^{|D|} m_{i,d} m_{i,e} \right] + \\ O \sum_{i=1}^{N} \sum_{j \neq i}^{N} n_i n_j \sum_{d=1}^{|D|} \left( m_{i,d} m_{j,d} - \sum_{e \neq d}^{|D|} m_{i,d} m_{j,e} \right) \therefore \end{aligned} \tag{32}$$

$$h = -C(|D| - 2) \land J_{idje} = \begin{cases} -C, & \text{if } i = j \land d \neq e \\ 2On_i n_j, & \text{if } i \neq j \land d \neq e \\ -2On_i n_j, & \text{if } i \neq j \land d = e \\ 0, & \text{otherwise.} \end{cases}$$

The ratio between the constraint scaling constant, $C$, and the objective scaling constant, $O$, selects how strongly the system should prioritize the one-hot encoding constraint compared to the objective metric of maintaining similar sums between partitions. An insufficient value of $C/O$ will allow resultant states of the PIM with high objective metrics, but violating a constraint, to have lower energies than the true solution, while an excessive value will prevent the PIM from efficiently exploring the state space.

The $|D|$-partition problem can be implemented without additional constraints by isotropic p-dits of the same dimensionality as the number of partitions. This neatly translates to the same $h$ vector and $J'$ matrix as for p-bits with 2-partition described in Equation (31). This is the case for many problems, as a p-bit can be thought of as an isotropic p-dit[2].

## *Integer Programming*

Solving ILP problems involves finding the optimum assignment of some column vector $\vec{x} = [x_1, x_2, ..., x_R]$ that maximizes

$$\vec{c} \cdot \vec{x}, \tag{33}$$

with $\vec{c}$ being a row vector of length $R$, while also being subject to a total of $Q$ constraints, organized in a matrix $S$ of dimension $Q \times R$, such that

$$S \cdot \vec{x} = \vec{b}. \tag{34}$$

The ILP Hamiltonian becomes

$$E = E_C + E_O = C \sum_{j=1}^{Q} \left( b_j - \sum_{i=1}^{R} S_{ji} x_i \right)^2 - O \sum_{i=1}^{R} c_i x_i. \tag{35}$$

The positive-valued constraint and objective constants, $C$ and $O$ respectively, are chosen to balance the influence of the constraints compared to the objective in exploring the problem's solution space[45].

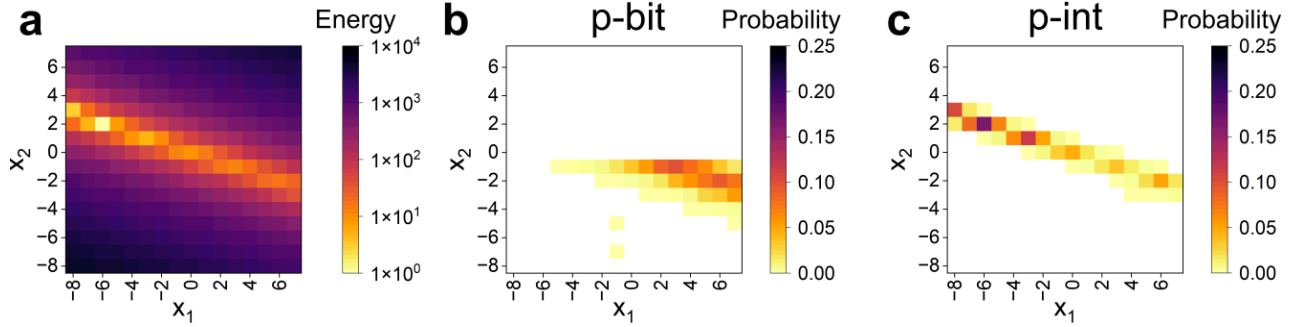

**Figure 3**: **A simple 2-variable ILP problem. A** The energy of all integer pairs for the ILP problem with the constraint $x_1 + 3x_2 = 0$, objective expression of $-x_1 + x_2$, $C = 1$, and $O = \frac{1}{5}$. The values have been linearly normalized so the ground state has an energy of 1. **B** A p-bit implementation of the same problem with $\beta = 0.1$. Even over 10,000 iterations, the PIM is unable to find its way to the top side of the ideal distribution due to the difficult transition from $-1 : 1111$ to $0 : 0000$. **C** The same problem implemented using the same parameters with p-ints. In contrast to the p-bit PIM, the p-int one is able to explore the entirety of the solution space.

Implementation of ILP using p-bits requires the substitution of each variable in Equation (35) as a sum of new two's-complement weighted variables[48]. As such, each variable requires $\lceil \log_2(B - A) \rceil$ p-bits to implement. In contrast, a single p-int can be used to represent each variable, which has apparent benefits for ILP problems with variables that have wide bounds; a situation that is common for useful ILP problems. More subtly, they allow a variable to increment or decrement smoothly along the number line towards its energy-minimizing value, rather than having to backtrack by flipping certain p-bits in the opposite direction. For instance, if six p-bits represent a variable with an optimum of 32, but a current value of 31, any single transition would be unfavorable, and a lower energy value achieved only after all six had flipped. By contrast, a p-int representing the same variable would need just one energy-favorable update. This results in p-int PIMs exploring the solution space more efficiently than p-bit PIMs, as apparent in a two-variable toy example whose true energy distribution is shown in Fig. 3a. As seen by its visitation distribution in Fig. 3b, the corresponding p-bit system is unable to transition between the $x_2 =$ $-1$ (1111 in binary) and the $x_2 = 0$ (0000 in binary) state over 10,000 iterations, defined in this work as a period in which a *single* p-dit may update. While over a sufficiently long trial would sample the ideal energy landscape, this example showcases the effect of large energy barriers introduced through binary encoding on exploration. The p-int system does not have this issue, and its distribution, shown in Fig. 3c, covers all low-energy regions of the true energy distribution of the problem.

For both implementations, an expansion of the respective versions of Equation (35) implies

$$h = -2C \sum_{j=1}^{N} \left(-b \cdot S_j^T\right) + O\vec{c} \ \land \ J = -\textbf{Hessian}(E) = -2C \sum_{j=1}^{N} S_j \cdot S_j^T. \tag{36}$$

However, for the p-bit PIM case, the identity elements are made zero. This can be done as $m_i \cdot m_i$ will always equal $+1$ for p-bits, and only the energy difference between states affects the probability.

P-ints can also be used to represent variables within IQP problems. They are implemented similarly to integer linear programming problems, however, the objective function is implemented using both the $h$ vector, to represent the linear terms, and the $J$ matrix, to represent the quadratic terms.

Inequality Constraints

Most integer linear programming problems include constraints that are represented using inequalities. To represent these within an Ising Hamiltonian, typically slack variables are introduced[55–57]. For instance,

$$x_1 - 5x_2 \leq 1 \tag{37}$$

becomes

$$x_1 - 5x_2 + x_3 = 1 \tag{38}$$

$$x_3 \geq 0 \Rightarrow A_3 = 0 \land B_3 \gg 0.$$

However, slack variables can introduce large energy barriers between configurations. For instance, with the above constraints the transition between the configurations $x = [00]$ and $x = [01]$ should be energy neutral. However, with the slack formulation, the transition from state $x = [001]$ to $x = [011]$ is highly energetically unfavorable, deviating far from the line specified in Equation (38). To maintain fulfillment of the constraint, the slack variable would first have to make the energy-unfavorable transition through each of the states between 1 and 6 to arrive at $x = [016]$. To make things worse, these introduced energy barriers vary in size depending on the magnitude of the coefficients in the original constraint. As such, the temperature cannot necessarily be adjusted to evenly flatten them.

Violation Variables

To mitigate these challenges, an alternative formulation of inequality constraints can be used with what we term violation variables. Violation variables hold a value of 0 if a constraint is satisfied, and $-1$ if it is violated, which can be realized using a p-int. Their $J$ matrix and $h$ vector connections are constructed from an equivalent slack variable representation. Firstly, the constraints' contributions to the $J$ matrix connections between traditional variables are zeroed, along with their $J$ matrix self-connections and $h$ vector offsets. Secondly, the $J$ matrix connections *from* the slack variables *to* the traditional variables are negated, while the $J$ matrix connections *from* the traditional variables *to* the slack variables remain the same. This results in a condensed but asymmetric $J$ matrix. An example is shown in Supplementary Note 7.

As violation variables have a value of 0 while their corresponding constraint is satisfied, they only affect traditional variables while a constraint is violated. Additionally, the correctional input to traditional variables does not scale with the degree of violation. While it seems like the inability for the system to prioritize fixing constraints that are majorly violated over those that are only minorly violated would be a disadvantage, this prevents constraints from distorting the energy landscape and thus creating hard-to-escape local minima.

There are two additional advantages of using violation variables: While conventionally, verifying that a seemingly stable state meets all inequality constraints requires manual computation, with this approach one can simply check to see that all violation variables remain 0 for some time. A further advantage is that, due to the lack of direct iteration between problem variables, a large portion of the $J$ matrix is known to have values of 0 and does not need to be stored in memory, resulting in significantly smaller storage space for large problems.

Scaled Sampling

Unlike with p-bits, there is sometimes a significant difference in total range between different p-ints in the same problem formulation. Intuitively, this leads to p-ints with smaller ranges being oversampled compared to those with large ranges. One way to adjust for this is by sampling each p-bit proportionally to its range, to approximately equalize the time it takes to travel from one end of each p-int's range to the other in a heavily weighted random walk[58].

**Results**

*|D|-Partition*

The more natural representation of isotropic p-dit PIMs compared to p-bit PIMs leads to better results on the same problem. For a small problem of 14 numbers, a p-bit PIM produces valid solutions (i.e., exactly one p-bit in each number's corresponding group is in the $+1$ state) for a reasonable range of constraint constants and temperatures, as shown in Fig. 4a and Fig. 4b. However, as the constraint constant rises, the quality of these valid solutions falls rapidly after a local minimum— primarily from the PIM's decreased emphasis on solution quality at higher $C$ values. The isotropic p-dit PIM for the same problem does not face this trade-off and is able to have much better solution quality, with a comparison to the p-bit PIM near the mean best error local minimum shown in Fig. 4c. This is confirmed experimentally, with ASIC results closely mirroring simulation results for both the p-bit and isotropic p-dit$^3$ implementations. Considering non-zero datapoints, an approximate 34x improvement is seen in trials-to-solution of the ASIC isotropic p-dit$^3$ implementation compared to the ASIC p-bit implementation.

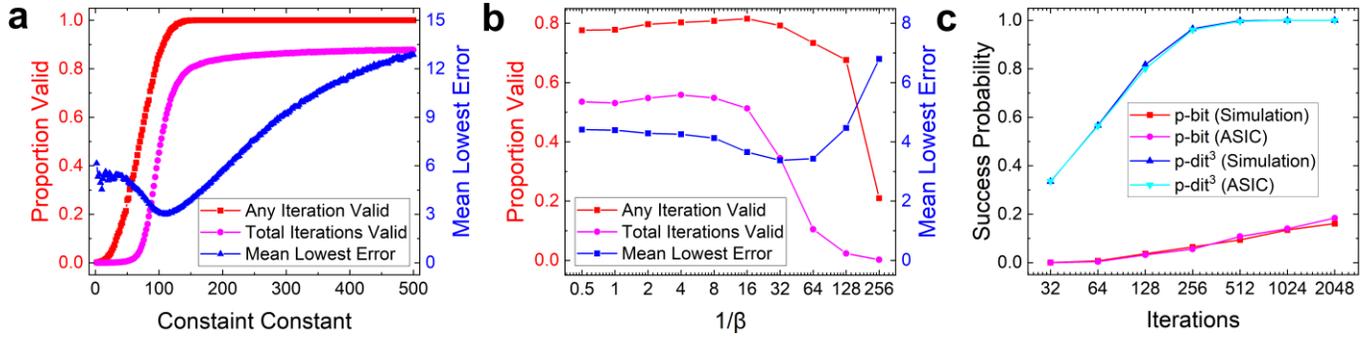

**Figure 4**: **3-partition problem. A** Results from a p-bit implementation of a 3-partition problem composed of 14 randomly generated integers, ranging between 1 and 6, over trials of 512 iterations with a constant $\beta$ of $\frac{1}{32}$. As $C$ is increased, an increasing proportion of the PIM's states during each trial are valid encodings. For valid encodings the error was calculated as the total absolute difference between each partition's sum and that of the true solution. The lowest error found in each trial increases with $C$ for most of its range after a minimum. Notably, for low $C$ values, few or no states were valid in 10,000 trials, resulting in a noisy or undefined mean error, respectively. $O$ is kept constant at 1. **B** The same setup, except that a constant $C$ of 94 is used and $\beta$ is swept. Lower temperature correlates with a higher proportion of valid states; however the best error is minimized at a medium temperature. **C** The same problem, sweeping the number of iterations performed for a p-bit PIM with a $C$ of 94 and an isotropic p-dit$^3$ PIM, both at a $\beta$ of $\frac{1}{32}$. For low numbers of iterations, no trials resulted in the p-bit PIM finding the true ground state solution. For each number of iterations, the isotropic p-dit$^3$ system had a significantly higher success rate. In both cases, results from the ASIC over at least 250 trials closely matched those from simulations over 10,000 trials.

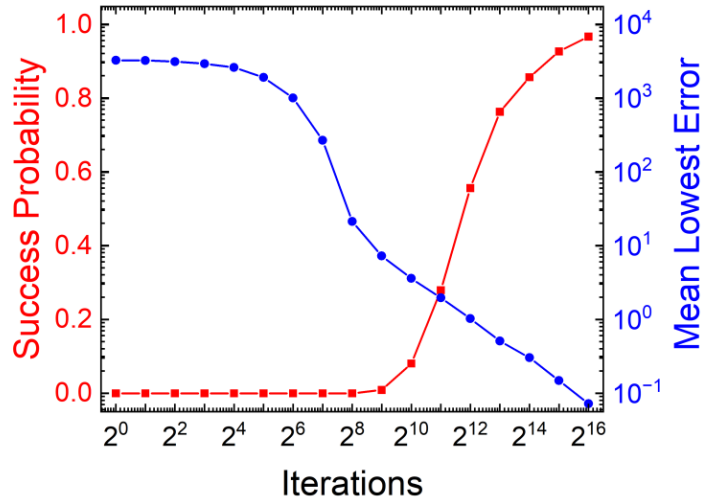

**Figure 5**: **6-partition problem.** Results from an implementation of a 6-partition problem of 1,000 randomly generated numbers between 1 and 100, using isotropic p-dits[6]. The error reduces as the number of iterations increases, with a sharp decrease starting at around $2^7$ iterations. When the error is sufficiently low, beginning around $2^{10}$ iterations, the true solution success rate grows rapidly. A linear $\beta$ sweep from $\frac{1}{32}$ to 1 is used over 1,000 trials for each data point.

Additionally, a larger problem of 1,000 randomly generated numbers between 1 and 100 was investigated with isotropic p-dits[6]. A rapidly decreasing error, starting when trial length exceeds around $2^7$ iterations, and rapidly increasing probability of finding the true ground solution, starting when trial lengths exceed around $2^{10}$ iterations, are shown in Fig. 5. For sufficiently long trial lengths, the ground solution was found in nearly every trial.

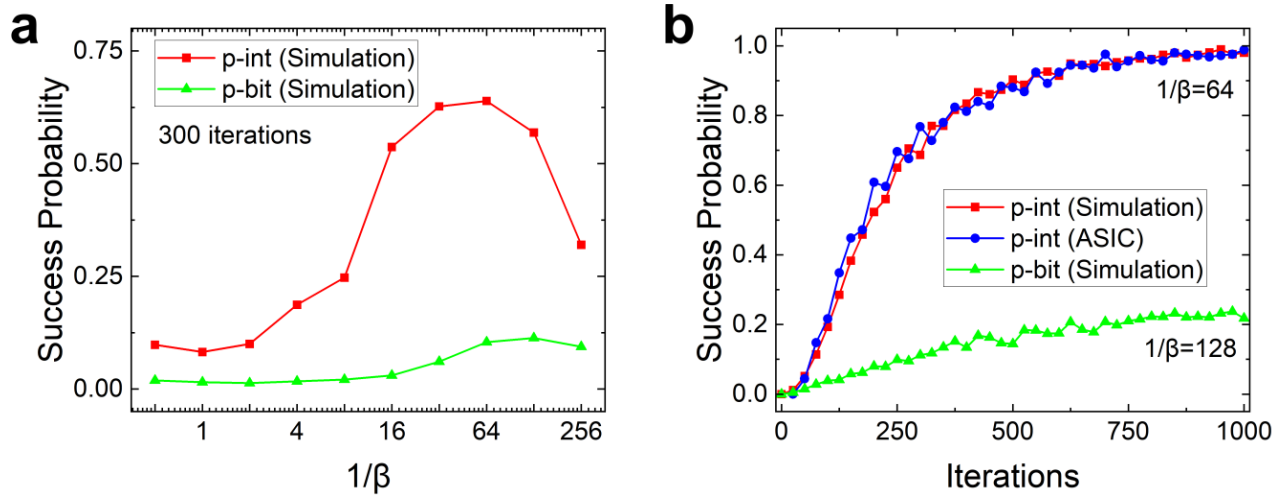

**Figure 6**: **Change-Making Problem. A** The proportion of 300 iteration-long trials using variable temperatures in which the true solution to a change-making problem was found. Specifically, the problem was composed to find change for a sum of $1.34 using the minimum number of coins worth 3¢, 4¢, 7¢, and 11¢, respectively. A $C$ of 1 and $O$ of 96 were used. **B** Simulated and ASIC results showing the p-int implementation greatly outperforming the p-bit implementation, with both using the best $\beta$ found from the first plot ($\frac{1}{64}$ for p-int and $\frac{1}{128}$ for p-bit). Each simulated data point is averaged over 1,000 trials, while each experimental data point is averaged over 250 trials.

*Integer Programming*

An ILP representation of the change-making problem, in which a minimum number of coins is sought whose value sum to a selected currency amount, is explored in Fig. 6. While the p-bit and p-int implementations use the same constraints and objective, they have slightly shifted optimal constant temperatures, $\beta_{opt}$, as shown by Fig. 6a with regards to finding the ground state solution of 14 coins. These $\beta_{opt}$ values were then used in a series of trials of variable length. These simulated results for p-bits, and both simulated and experimental results for p-ints, are shown in Fig. 6b. The simulated and ASIC p-int implementation greatly outperforms the simulated p-bit implementation across the entire trial length range. Averaged over all non-zero datapoints, an approximate 5.3x improvement is seen in trials-to-solution of the ASIC p-int implementation compared to the p-bit implementation.

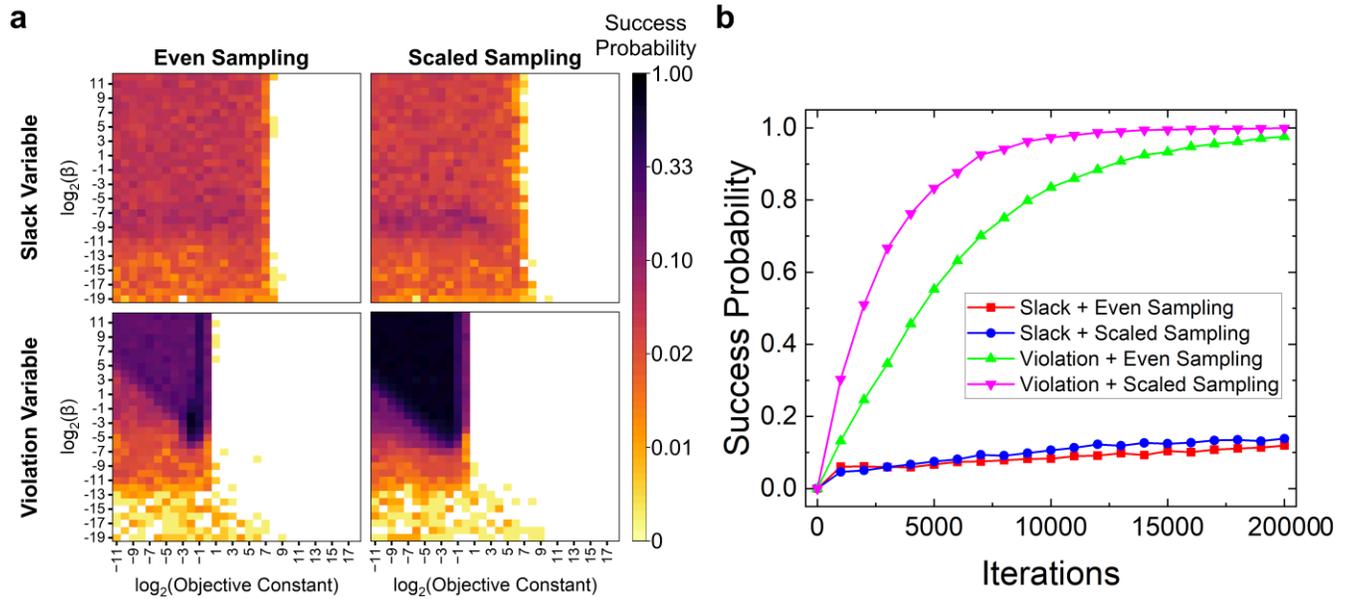

**Figure 7**: **Fixed-Charge Problem. A** Heat maps of the proportion of 500 trials of 8,192 iterations in which the optimum solution was found in a fixed-charge problem. During each trial, $\beta$ was swept from the listed $\beta_0$ value to 32 times its value, using a base-2 logarithmic sweep. $C$ was kept constant at 1. In the top left, a traditional slack variable representation of the problem with even sampling was used. A relatively wide range of both $\beta_0$ and $O$ resulted in decent performance. In the top right, a slack variable representation and variable range scaled sampling were used. Compared to the even sampling, the optimal range is slightly more defined. In the bottom left, a violation variable representation with an even sampling as used. Good performance is found for a large region of $\beta_0$ and $O$. In the bottom right, a violation variable representation and scaled sampling were used. Great performance is found for a large region of the explored variables. **B** Performance of the four representations for the same problem over 1,000 trials of varying length. A pronounced increase in performance is found by using a violation variable representation over traditional slack ones. A clear increase is found by using adjusted sampling over even sampling in the violation variable case, and a lesser but still distinguishable increase is shown after 5,000 iteration trials in the slack variable case. For both slack representation formulations, $\beta_0 = \frac{1}{256}$ and $O = 2$, while both violation representation formulations used $\beta_0 = \frac{1}{4}$ and $O = \frac{1}{4}$.

Fig. 7 shows four approaches to a slightly modified fixed-charge ILP problem[55] representing a hypothetical business's choice of which clothing items to manufacture, which is traditionally represented using six variables, 14 constraints, and an objective function. For even and scaled sampling with both a traditional slack representation of the problem and a violation variable representation, a parameter sweep over the objective constant and $\beta_0$ was performed. During each trial, $\beta$ was swept from $\beta_0$ to $32\beta_0$ using a base-2 scaling. From Fig. 7a, it was found that for both sampling methods, the slack formulation had a parameter optimum near $\beta_0 = \frac{1}{256}$ and $O = 2$, while both violation representations had optima near $\beta_0 = \frac{1}{4}$ and $O = \frac{1}{4}$. These values were then used to explore the behavior of the four selected formulations over trials of various lengths. As Fig. 7b shows, the violation formulation offers significant improvement in finding the absolute solution compared to the traditional slack formulation. Additionally, the selected scaled sampling procedure offers major improvements for the violation formulation over the entire range, and minor improvement for the slack formulation for trial lengths of more than ~5,000 iterations. Averaged across non-zero datapoints, an ~10x improvement in trials-to-solution is seen by the violation formulation using scaled sampling compared to the traditional slack formulation using even sampling.

Integer Quadratic Programming

A non-convex, 17-variable problem with 51 linear constraints, of which 34 were bounds, and a quadratic objective function were randomly generated. To track the value of the custom objective function, a capability not included within the ASIC design, and to allow for parallel instances of the problem, an FPGA implementation was used. The design included two cores, each of which contained two pipelined instances of the problem. As the system clock of the FPGA was 50 MHz, this was equivalent to 4 parallel instances which updated at 25 MHz. During each trial, the initial state of each instance had all variables set to a value of 0 to avoid the need for preprocessing. As the trial progressed, each instance searched for the true solution independently. If no instance found the solution within a reasonable time, each instance was reset at a pre-selected cutoff iteration, without resetting the trial's timer. As the distribution of iterations-to-solution necessarily has a long right tail, smaller cutoff values were shown to be preferable.

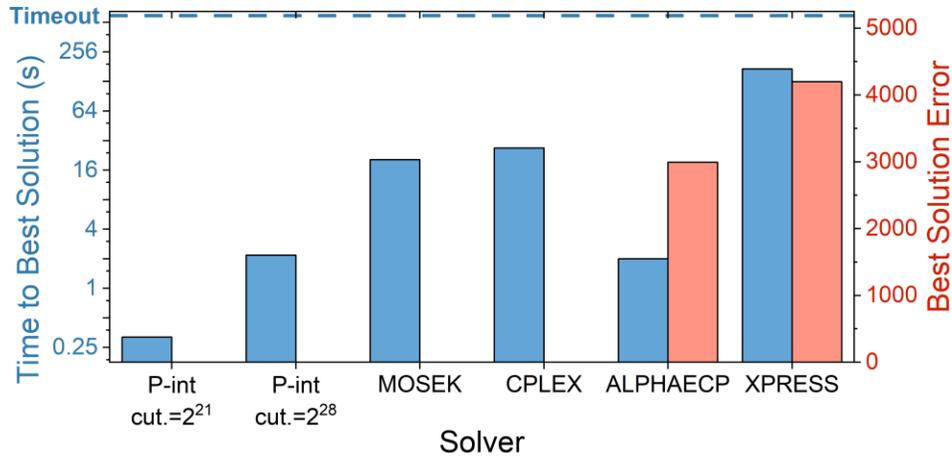

**Figure 8**: **Non-convex IQP Problem Comparison.** A comparison of the time required to produce a solution, and the quality of solutions for an IQP problem for each solver. The generated problem includes 17 integer variables, 51 linear inequality constraints (34 being bounds), and a quadratic objective function. The p-int implementation was synthetized on an FPGA. It contained two cores with two pipelined instances each and used $\beta_0 = 3.3$ and $O = 0.0008$. If no solution was found by the iteration cutoff, the state of the system was reset, with the trial continuing. The remainder are state-of-the-art software solvers included in GAMS. The p-int data is composed from 25 trials, while the solvers data arose from a trial with the default seed. Despite the slower 50 MHz clock speed of the FPGA compared to the 4.31 GHz boost of the CPU, the p-int system in both configurations outperformed all software solvers which found the problem's true solution. Configured with a cutoff of $2^{21}$, a ~64x improvement in time-to-solution was observed for the p-int solver compared to the MOSEK solver.

The performance of this system with two cutoff values was compared to several state-of-the-art software solvers parallelized across 4 threads, with the results shown in Fig. 8. The solvers considered were the standalone ones included in GAMS that were advertised as solving mixed-integer quadratic programming (MIQCP) problems and that were accessible with an academic license. Due to the complicated nature of the problem, two of the four solvers were unable to find the ground solution within the allotted 10-minute limit: the error of their best-found solution is plotted on the right y-axis of Fig. 8.

The p-int system with both a large ($2^{28}$) and small ($2^{21}$) cutoff vastly outperformed even the best state-of-the-art solver, by ~10x and ~64x respectively, in finding the true solution. The performance of the p-int system is especially impressive considering the higher 4.31 GHz boost clock of the CPU.

## Discussion and Conclusions

| Problem | Complexity | Implementation Used | Required p-bits | Required p-dits | Platform | Performance Improvement |
|---|---|---|---|---|---|---|
| **Change-making ILP** | 8 bound constraints + 1 equality constraint + 1 objective | p-bits & p-ints | 16 | 4 | Software & ASIC | 5.3x p-int PIM vs. p-bit PIM |
| **3-partion problem** | 14 values | p-bits & isotropic p-dits[3] | 42 | 14 | Software & ASIC | 34x p-dit PIM vs. p-bit PIM |
| **6-partition problem** | 1,000 values | isotropic p-dits[6] | 6,000 | 1,000 | Software | - |
| **Fixed-charge ILP** | 12 bound constraints + 5 inequality constraints + 1 objective | p-ints (each combination of slack/violation + even/scaled) | 55 | 11 | Software | 10x violation + scaled vs. slack + even |
| **Non-convex IQP** | 34 bound constraints + 17 inequality constraints + 1 objective | p-ints (violation + even sampling) | 289 | 34 | FPGA | 64x p-int PIM vs. best state-of-the-art software solver |

**Table 1**: **Summary of experiments.** A summary of the benchmarks implemented in this work. The table summarizes the type of problem, its complexity, the probabilistic architecture used to solve it, the size of a p-bit PIM needed to solve it (whether p-bits were used or not), the size of the p-int or isotropic p-dit PIM needed to solve it, the platform, and the observed performance improvement in trials-to-solution or time-to-solution (when applicable) compared to a stated reference.

In this work, we have described a generalized mathematical spin model which supports continuous length components along an arbitrary count of dimensions. We further formalized the probabilistic d-dimensional bit which stochastically oscillates between discrete states within this space. As a restriction of the general p-dit definition, the concept of isotropic p-dits was introduced. A single isotropic p-dit can be used to represent the value of a categorical variable that is traditionally represented with a group of p-bits. While a one-hot encoding using p-bits results in a PIM where most configurations are invalid, an isotropic p-dit implementation will never exist in an invalid configuration. This advantage increases with the complexity of the problem, as the ratio of invalid to valid assignments grows exponentially with the number of various represented. For p-bit implementations, the one-hot constraint must be implemented within the $J$ matrix and $h$ vector alongside any other constraints or objectives of the problem. This necessitates careful tuning of their relative strength, which adds another non-trivial aspect to achieving optimization. If the one-hot encoding constraint is enforced too weakly, the PIM might converge to an invalid configuration where it is violated but the objective has a favorable value. On the other hand, if the one-hot encoding is enforced over-zealously, then the system will have a difficult time leaving a local minimum, as it takes at least two flips to move from one valid one-hot encoding to another. We have demonstrated experimentally, for a small 3-partition problem, a strong ~34x improvement of an isotropic p-dit[3] PIM compared to a p-bit PIM, even after finding the optimum relative strength of the one-hot encoding for the p-bit implementation. While the $N$-partition problem is a natural demonstration of the power of the isotropic p-dit formulation, which is also why a 1,000-node 6-partition instance was further demonstrated as solvable, we believe many other problems can be attacked in this manner.

We further introduced another useful restriction on the definition of p-dits, termed probabilistic integers, which allow for a more natural representation of numeric values with regular, discrete steps. The most promising application of p-int PIMs is to solve integer programming problems: the common language in which many real-world optimization problems are expressed. While p-bits can represent integer values in these problems, the necessary binary encoding introduces energy barriers between adjacent states: a challenge not encountered with p-ints. We have demonstrated a large performance improvement of a p-int PIM compared to a p-bit PIM for an ILP problem (~5.3x) with an ordinary Ising representation. This result, along with the others of this work, are summarized in Table 1.

It is worth noting, however, that many practical problems are complicated, composed of a mixture of equality, and inequality constraints, of small and large range variables, and of linear and quadradic objectives. We thus believe that there is much work to be done in maturing the field of p-dit-based Ising solvers to allow for their widespread adaptation for this task. We have introduced the concept of violation variables as a method of representing inequality constraints within an Ising Hamiltonian. By breaking the symmetry of the coupling matrix, they allow a PIM to more smoothly explore a problem's energy landscape compared to the traditional slack variable method. We further explored sampling p-ints with different frequencies depending on the range they represent. Finally, we implement a quadradic objective function across both the $h$ vector and the $J$ matrix. Using an FPGA with a 50 MHz clock to implement a p-int PIM, we were able to demonstrate a massive, ~64x improvement in time-to-solution compared to the best state-of-the-art software solvers tested, for an IQP problem. This demonstrates the potential for probabilistic computing methods to replace existing integer programming solvers.

An important aspect of the practicality of the proposed p-dits is the ability to implement them in hardware. To showcase the ability to create CMOS-based p-dit-based PIMs, we designed, fabricated, and verified a digital ASIC test chip. Crucially, it was able to perform a full iteration of the PIM within two clock cycles. This suggests that PIM ASICs of all three p-dit types shown here can be designed to achieve ultra-fast update rates.

## Methods

### ASIC design

The ASIC PIM was defined using RTL Verilog code which was processed by the OpenLane RTL to GDSII pipeline, using the Skywater 130 nm open-source process design kit (PDK). The PIM was manufactured using the Efabless multi-project wafer (MPW) service, which also provided the design for a co-integrated small CPU based on a VexRiscv minimal+debug configuration[59]. An oscillator running at 10 MHz was used as the ASIC's clock for all experiments shown.

### Experimental code

All non-trivial problems and their Ising representations were created using self-made Python 3 code. All simulated data were gathered using self-made C++ code. Self-made C code executed on the RISC-V CPU was used to run all experimental trials. Communication with the RISC-V CPU, to upload trial code and to record results, was done using the UART protocol over a USB cable.

### IQP Comparison

Solvers used for the IQP comparison were bundled with GAMS 49.3.0[60] and accessed using GAMS Studio 1.20.2. To the best of our knowledge, all local, standalone (not utilizing other solvers included in GAMS as subsolvers) non-convex MIQCP solvers that were included in an academic license have been considered. For each trial, the solvers were instructed to use 4 threads, and to have a timeout of 10 minutes, with all other parameters kept at their default values, including the seed. For CPLEX, 'OptimalityTarget' was set to '3' to allow it to process a non-convex problem. The solution time for each solver was extracted from the produced logfiles, where it is reported differently for each solver. CPLEX and MOSEK reported solution time to the nearest hundredth of the second, while ALPHAECP and XPRESS reported solution time to the nearest second. Total execution time was not considered. All solvers finished execution on their own apart from XPRESS which was halted after the timeout period. This comparison was conducted on a Ryzen 7 4700U which has a frequency of 2.0 GHz and a Turbo Clock of 4.1 GHz. Memory utilization was not a limiting factor for any solver.

*FPGA design*

All FPGA demonstrations were conducted using a Terasic Cyclone IV Altera DE2-115. A 50 MHz oscillator included on the FPGA was used as the driving clock. The design was defined by RTL SystemVerilog code compiled by Quartus Prime Version 23.1std.0 Build 991. One push-button was used to reset the pseudo-random number generator to an initial seed. The other independently reset the state of the PIM. The pseudo-random number generator is based on the state update procedure of the 32-bit PCG pseudo-random number generator[61]. It has a period of $2^{33}$ states, corresponding to ~86 seconds. For each set of initial, randomly generated seeds, 5 trials were conducted. Readout of the iteration in which the true solution was found was done through LEDs.

29. Kalinin, K.P., Amo, A., Bloch, J. & Berloff, N.G. Polaritonic xy-ising machine. *Nanophotonics* **9**, 4127–4138 (2020).

30. Luo, S. *et al.* Classical spin chains mimicked by room-temperature polariton condensates. *Physical Review Applied* **13**, 044052 (2020).

31. Camsari, K. & Datta, S. Waiting for quantum computing? try probabilistic computing. *IEEE Spectrum* (2021).

32. Camsari, K. Y., Faria, R., Sutton, B. M. & Datta, S. Stochastic p-bits for invertible logic. *Physical Review X* **7**, 031014 (2017).

33. Liu, Y. *et al.* Time-division multiplexing ising computer using single stochastic magnetic tunneling junction. *IEEE Transactions on Electron Devices* **69**, 4700–4707 (2022).

34. Yin, J. *et al.* Scalable ising computer based on ultra-fast field-free spin orbit torque stochastic device with extreme 1-bit quantization. In *2022 International Electron Devices Meeting (IEDM)*, 36–1 (IEEE, 2022).

35. Lu, A. *et al.* Scalable in-memory clustered annealer with temporal noise of finfet for the travelling salesman problem. In *2022 International Electron Devices Meeting (IEDM)*, 22–5 (IEEE, 2022).

36. Kaiser, J. *et al.* Hardware-aware in situ learning based on stochastic magnetic tunnel junctions. *Physical Review Applied* **17**, 014016 (2022).

37. Camsari, K. Y., Sutton, B. M. & Datta, S. P-bits for probabilistic spin logic. *Applied Physics Reviews* **6** (2019).

38. Kaiser, J. & Datta, S. Probabilistic computing with p-bits. *Applied Physics Letters* **119** (2021).

39. Grimaldi, A. Probabilistic and oscillatory Ising machines for combinatorial optimization: from software to hardware-acceleration with spintronics. *Ph.D. thesis, Università degli Studi di Messina* (2023).

40. Singh, N. S. *et al.* Cmos plus stochastic nanomagnets enabling heterogeneous computers for probabilistic inference and learning. *Nature Communications* **15**, 2685 (2024).

41. Okuyama, T., Hayashi, M. & Yamaoka, M. An ising computer based on simulated quantum annealing by path integral monte carlo method. In *2017 IEEE international conference on rebooting computing (ICRC)*, 1–6 (IEEE, 2017).

42. Zhang, T. *et al.* A review of ising machines implemented in conventional and emerging technologies. *IEEE Transactions on Nanotechnology* (2024).

43. Zhang, T., Tao, Q., Liu, B. & Han, J. A review of simulation algorithms of classical ising machines for combinatorial optimization. In *2022 IEEE International Symposium on Circuits and Systems (ISCAS)*, 1877–1881 (IEEE, 2022).

44. Gyoten, H., Hiromoto, M., & Sato, T. Enhancing the solution quality of hardware ising-model solver via parallel tempering. In *2018 IEEE/ACM International Conference on Computer-Aided Design (ICCAD)* (pp. 1-8). (IEEE, 2018).
Page 36 of 38

**Acknowledgments:** This work was supported by the U.S. National Science Foundation (NSF) under award numbers 2322572, 2425538, and 2400463. A.G., D.V. and G.F. are members of the Petaspin team and acknowledge the support from Petaspin association ([www.petaspin.com](www.petaspin.com)). G. F. and A.G. acknowledge the support from the project PRIN 2020LWPKH7, "The Italian factory of micromagnetic modelling and spintronics". The work of D. V. thanks the project PE0000021, "Network 4 Energy Sustainable Transition – NEST", funded by the European Union – NextGenerationEU, under the National Recovery and Resilience Plan (NRRP), Mission 4 Component 2 Investment 1.3 - Call for tender No. 1561 of 11.10.2022 of Ministero dell'Università e della Ricerca (MUR) (CUP C93C22005230007).